\documentclass[prl,twocolumn,epsf,psfig]{revtex4}
\usepackage{graphicx}
\DeclareGraphicsExtensions{.pdf,.png,.gif,.jpg}
\usepackage{float}
\usepackage{subfig}
\usepackage{epsfig}
\usepackage{wrapfig}
\usepackage{bbold}
\usepackage{amsmath}
\usepackage{color}

\restylefloat{figure}

\begin{document}

\title{Two-band atomic superfluidity in the presence of an orbital Feshbach resonance}

\author{Andrew Vincent and Theja N. De Silva}
\affiliation{Department of Physics and Biophysics,
Augusta University, Augusta, Georgia 30912, USA.}

\begin{abstract}
We study static superfluid properties of alkali-earth-like Fermi atomic systems in the presence of orbital Feshbach resonance. Using a two-band description of the ground state and excited state and a mean-field approximation of the intra-band atomic pairing, we investigate the phase transitions and crossover between BCS and Bose-Einstein-condensate (BEC) superfluidity. At zero temperature, we find exact BCS analytical solutions for the mean-field gap equations and number equations. Using these exact solutions, we calculate various static properties, such as superfluid order parameters, chemical potentials, density variations, density profiles, correlation and coherence lengths, ground-state energy, and Tan's contact density across the entire BCS-BEC crossover region. We anticipate that our closed-form analytical results can be used as a benchmark for future experimental and theoretical investigations and will have an impact on the current understanding of two-band superconductors such as MgB$_2$.

\end{abstract}

\maketitle

\section{I. Introduction}

As a result of the high degree of independent tunabilty 0f, for example, interaction parameters, types of species, and number of hyperfine spin states and their population, cold-atomic systems are considered flexible quantum simulators for quantum few- and many-body physics~\cite{CA1, CA2, CA3,CA4, CA5, CA6}. The evolution from BCS to  Bose-Einstein-condensate (BEC) atomic pairing in single-band superfluids has attracted tremendous interest recently, both experimentally and theoretically~\cite{NW1, NW2, NW3, NW4, NW5, NW6, NW7, NW8, NW9, NW10, NW11, NW12}. This is due to the fact that both BCS and BEC superfluidities have strong roots in many areas of physics, including condensed matter, material science, atomic physics, astrophysics and quantum electrodynamics~\cite{BCS}. Condensation of alkali atoms such as $^6$Li and $^{40}$K shows BCS-BEC crossover upon tuning the energies of the two-body closed-channel bound state and two interacting open-channel scattering states. This energy tuning is done by applying an external magnetic field to create a Zeeman energy shift of atomic electronic spin states allowing for control of the single-band s-wave scattering length. This way of controlling the effective hyperfine interactions between atoms is called magnetic Feshbach resonance. Further, for the alkali-earth-like Fermi atoms, such as $^{173}$Yb, interactions cannot be controlled using the magnetic Feshbach resonance due to their insensitivity to the magnetic field owing to their zero-electronic angular momentum. However, as discussed below, the Zeeman energy shift in nuclear spin-states of alkali-earth-like atoms can be used to tune the coupling between closed and open channels and control interorbital state interactions known as orbital Feshbach resonance. Therefore, a single-band description and open-channel scattering are sufficient for the BCS-BEC evolution of magnetic Feshbach atomic systems. For orbital Feshbach atomic systems, the two-band description and both open- and closed-channel scattering is required. Thus, alkali-earth-like Fermi atomic systems have become a recent topic of interest for understanding two-band superfluidity and superconductivity. 

Soon after the prediction of orbital Feshbach resonances~\cite{MOD1},  two research groups independently identified two-body scattering resonances related to the orbital Feshbach resonance of the $^{173}$Yb atomic system. These observations ignited a significant amount of theoretical research studying the many-body phenomena associated with alkali-earth-like Fermi atomic systems and their connection to two-band superconductors~\cite{XFR1, XFR2, MOD2, MOD3, MOD4, MOD5, MOD6, MOD7, MOD8, MOD9, MOD10, MOD11, MOD12, MOD13}. By calculating the effective range and magnetic field dependence on the scattering length of alkaline-earth-metal-like $^{173}$Yb systems, Xu \textit{et al.} proposed to use narrow orbital Feshbach resonance (OFR) to achieve a critical temperature that is higher than that of the wide resonance~\cite{MOD2}. Their results have inspired both experimentalists and theorists to undertake a wealth of research activities related to OFR systems. At the zero-temperature limit of the OFR systems, using an L-J potential, He \textit{et al.} studied the stability, equation of state, and potential observation of the Leggett mode~\cite{MOD14}. They found dynamical instability through the density excitation spectrum of the system at larger momentum, thus supporting the existence of metastable conditions for experiments. However, that work found a severely damped Leggett mode due to the small singlet scattering length. Using zero-temperature mean-field theory, Iskin introduced two intraband order parameters and investigated the in-phase and out-of-phase solutions of the order parameters and found that OFR systems are exactly analogous to two-band  s-wave superconductors~\cite{MOD3}. This two-band superconductor analogy motivated Klimin \textit{et al.} to study Leggett collective modes in OFR-like systems using the Gaussian pair fluctuation propagator~\cite{MOD15}. Again, restricting their study to the zero temperature and interorbital pairing, Laird \textit{et al.} studied the impact of frustration in the few- and many-body limits~\cite{MOD6}. They found that the extra closed-channel atoms cause significant suppression of the ground-state binding energy; therefore, the pairing in the resonance channel of OFR systems is closer to the weaker BCS side of the resonance. Going beyond previous mean-field studies to include fluctuations, Kamihori \textit{et al.} studied the superfluid properties of OFR systems by using Nozieres and Schmitt-Rink type pair-fluctuation theory~\cite{MOD5}. This work investigated the band dependence on condensate fractions and superfluid collective modes and confirmed the absence of the Leggett mode predicted in previous studies. 

The aforementioned theoretical work motivated us to search for a simplified approach to the solution of BCS mean-field formalism and predict various static properties. Some of the work presented in this paper is somewhat complementary to the work already mentioned above~\cite{MOD5, MOD6}. However, we derive those complimentary results from exact analytical BCS solutions. One of the major achievements in our work is finding exact analytical BCS solutions to two-band OFR systems. Further, going beyond previous research, we use our analytical solutions to predict various  experimentally observable static properties. They include superfluid order parameters, chemical potentials, density variations, density profiles, correlation and coherence lengths, ground-state energy, and Tan's contact density across the entire BCS-BEC crossover region. As the real-space atomic density variations and Tan's contact density are easily accessible in experiments and they are closely related to all microscopic details of the system, our prediction will have an impact on current and future OFR experiments.

This paper is organized as follows. In Sec. II, we discuss the differences and similarities between magnetic Feshbach resonance and orbital Feshbach resonance and introduce the two-orbital bands for the orbital Feshbach atoms. In Sec. III, we introduce our two-orbital-band model Hamiltonian with interorbital and intraorbital interaction parameters. Section IV is dedicated to the mean-field approximation. We use this approximation to bring our Hamiltonian into diagonal form, which allows us to find the grand potential of the atomic system. Using the grand potential, we derive our self-consistent equations for both number equations and gap equations in Se. V. In Sec. VI, by restricting ourselves to zero temperature, we present our exact analytical BCS solutions to the self-consistent equations. In this section, we present all our results for static properties. Finally, in Sec. VIII, we summarize our results.

\section{II. Shape resonance, magnetic Feshbach resonance, and orbital Feshbach resonance}

In atomic physics, resonance is the emergence of stable bound states from scattering states of two atoms or vice versa. When this happens, two atoms interact very strongly, and their scattering cross-section becomes very large~\cite{CA3}. A shape resonance occurs when the inter-atomic potential has a weakly bound state which lies close to the continuum threshold of the atoms within the same potential. Since the bound state and the scattering states are in the same potential, it is extremely difficult to experimentally control the interaction properties of the atoms. A magnetic Feshbach resonance occurs when the interaction between two atoms has potentials depending on whether the total spin is a singlet or a triplet. When the two atoms scatter in one potential and the other potential supports a stable bound state, one can utilize the Zeeman energy by applying an external magnetic field to control the scattering properties. This is possible because the magnetic moment of the colliding atoms is different from that of bound states. For example, in cold gas experiments with alkali atoms such as $^6$Li and $^{40}$K, the scattering between atoms can be characterized by a single parameter, the s-wave scattering length $a$~\cite{NW1, NW2, NW3, NW4}. This is due to the fact that at characteristic densities and ultra-cold temperatures, only isotropic and short-range s-wave scattering between atoms can take place. Using the magnetic Feshbach resonance, the s-wave scattering length can be tuned from small positive values to small negative values through positive and negative infinities.  In the attractive interaction regime where $a < 0$, the atoms form BCS pairs such that the ground state is a BCS superfluid. In the repulsive interaction regime where $a > 0$, the atomic potential supports a two-body molecular bound state in vacuum such that the ground state is a BEC of these molecules. In between these two ground states, there is a smooth crossover where $a$ changes its sign as it passes through $\infty$. 

For alkali-earth-like atoms such as $^{173}$Yb, the magnetic Feshbach resonance is absent owing to their total zero electronic spin originating from the completely filled outer shells in the ground state~\cite{XFR1, XFR2}. However, the scattering properties of alkali-earth-like atoms can be tuned by using a mixture of atomic gas in the long-lived excited state $^3$P$_0$ and the ground state $^1$S$_0$. The excited state $^3$P$_0$ is a state where one electron is excited to a $p$ orbital and the two valence electrons form a spin triplet. The life time of this excited state is long because the dipole transition to the ground state $^1$S$_0$ is spin-forbidden. In addition to the nuclear spin, these two orbital states, $^3$P$_0$ and $^1$S$_0$, can be utilized to induce Feshbach resonance between two atoms. The coupling between resonance channels through inter-orbital nuclear-spin exchange interactions is called orbital Feshbach resonance.

\section{III. Model Hamiltonian}

As the model Hamiltonian, we use the same Hamiltonian as in Refs.~\cite{MOD3, MOD5, MOD6}. We consider two nuclear spin states denoted by the spin $\sigma = \uparrow, \downarrow$, in two energy states $^3$P$_0 = |e \rangle$ and $^1$S$_0 = |g \rangle$. For $^{173}$Yb atoms, $\sigma$ can be two of the six hyper-fine states of nuclear spin $I = 5/2$ states. The energy states $|e \rangle$ and $|g \rangle$ are regarded as two orbital states $\alpha$. The basis states can the be defined as the ket $|\alpha, \sigma \rangle$. In the presence of an external magnetic field, orbital and spin degrees of freedom are coupled, and two-body internal states are defined as open and closed channels~\cite{MOD1}. The open channel $|o\rangle$ is designated by the antisymmetric state with one atom in state $|g, \downarrow \rangle$ and the other atom in state $|e, \uparrow \rangle$. Similarly, the closed channel $|c\rangle$ is designated by the state with one atom in state $|g, \uparrow \rangle$ and the other atom in state $|e, \downarrow \rangle$,

\begin{eqnarray}
|\alpha = 1 \rangle \equiv |o\rangle = \frac{1}{\sqrt{2}}(|g, \downarrow; e \uparrow \rangle - |e, \uparrow; g \downarrow) \rangle \\ \nonumber 
|\alpha = 2 \rangle \equiv |c\rangle = \frac{1}{\sqrt{2}}(|g, \uparrow; e \downarrow \rangle - |e, \downarrow; g \uparrow \rangle).
\end{eqnarray}

Both channels have the same threshold energies in the absence of a magnetic field $B$. However, in the presence of a magnetic field, these two states split by an energy amount $\delta \equiv \delta(B)$, which can be incorporated through chemical potentials for the atoms in the open channel and closed channel, $\mu_{\sigma o} = \mu_{\sigma}$ and $\mu_{\sigma c} = \mu_{\sigma} - \delta/2$, respectively. With the application of an external magnetic field $B$, the interchannel separation $\delta$ can be tuned to be below or above the binding energy of a molecule in the closed channel. This allows one to tune the scattering length of the atoms in the open channel from negative values to positive values. The scattering length diverges when the interchannel separation $\delta$ exactly matches the binding energy.

The model Hamiltonian $H = H_0 + H_I$ is the sum of kinetic energy and interaction energy between atoms~\cite{MOD2, MOD3}. First, we define a fermionic creation/annihilation operator $\psi^\dagger_{\sigma \alpha}(\vec{r}_i)/\psi_{\sigma \alpha}(\vec{r}_i)$ to represent the creation/annihilation of an atom at position $\vec{r}_i$, with spin $\sigma$ in orbital $\alpha$, where we take $\alpha =1$ as the open channel and $\alpha =2$ as the closed channel. The kinetic energy term is then written as, 

\begin{eqnarray}
H_0 = \sum_{\sigma \alpha} \int d^3\vec{r}_i \psi^\dagger_{\sigma \alpha}(\vec{r}_i) \biggr(-\frac{\hbar^2 \nabla^2}{2M} - \mu_{\sigma \alpha}  \biggr)\psi_{\sigma \alpha}(\vec{r}_i), 
\end{eqnarray}

\noindent where $h = 2 \pi \hbar$ is Planck's constant and $M$ is the mass of an atom. In the two-band description, the interaction energy can be written as

\begin{eqnarray}
H_I = \sum_{\alpha \beta} \int d^3\vec{r}_i d^3\vec{r}_j \psi^\dagger_{\uparrow \alpha}(\vec{r}_i) \psi^\dagger_{\downarrow \alpha}(\vec{r}_i) \\ \nonumber  \times V_{\alpha \beta}(|\vec{r}_i-\vec{r}_j|)  \psi_{\downarrow \beta}(\vec{r}_j) \psi_{\uparrow \beta}(\vec{r}_j),
\end{eqnarray}

\noindent where $V_{\alpha \beta}(|\vec{r}_i-\vec{r}_j) \equiv V_{\alpha \beta} \delta(\vec{r}_i - \vec{r}_j)$ is the interaction potential between two atoms at positions $\vec{r}_i$ and $\vec{r}_j$ in orbitals $\alpha$ and $\beta$. This $\delta$-function interaction potential is reasonable for cold-atom experiments as the range of the interaction potential is much smaller than the average inter particle distance for dilute atomic mixtures. Due to this relatively large inter particle distance, when atoms are interacting, they cannot resolve the detailed structure of the inter atomic potential. The interaction potential is related to the singlet scattering potential and triplet scattering potential as,  

\begin{eqnarray}
V_{oo} = V_{cc} = \frac{4 \pi \hbar^2 (a_{-} + a_{+})}{M} \equiv \frac{4 \pi \hbar^2 a_{so}}{M} \\ \nonumber
V_{oc} = V_{co} = \frac{4 \pi \hbar^2 (a_{-} - a_{+})}{M} \equiv \frac{4 \pi \hbar^2 a_{s1}}{M},
\end{eqnarray}

\noindent where $a_{-}$ and $a_{+}$ are the singlet and triplet scattering lengths, respectively.

\section{IV. Mean-field theory and thermodynamic potential}

In order to study the superfluidity and the static properties, we convert the interacting atomic Hamiltonian into an effectively non-interacting quasi-particle Hamiltonian using mean-field theory. The existence of superfluidity is determined by the uniform superfluid order parameter, $\Delta_{\alpha} = -\sum_\beta V_{\alpha \beta} \langle \psi_{\downarrow \beta}(\vec{r}_j) \psi_{\uparrow \beta}(\vec{r}_j) \rangle$, where $\langle O \rangle$ is the thermal or quantum mechanical expectation value of the operator $O$ with respect to the quantum mechanical system wave function. Using our mean-field theory, the interaction part of the Hamiltonian can be written as

\begin{eqnarray}
H_I = \sum_{\beta} \int d^3\vec{r} \Delta^{\ast}_\beta(\vec{r}) \psi_{\downarrow \beta}(\vec{r}) \psi_{\uparrow \beta}(\vec{r}) \\ \nonumber + \sum_{\alpha} \int d^3\vec{r} \Delta_\alpha(\vec{r}) \psi^\dagger_{\uparrow \alpha}(\vec{r}) \psi^\dagger_{\downarrow \alpha}(\vec{r}) \\ \nonumber - \sum_{\alpha \beta} \int d^3\vec{r} V_{\alpha \beta} \langle \psi^\dagger_{\uparrow \alpha}(\vec{r}) \psi^\dagger_{\downarrow \alpha}(\vec{r}) \rangle \langle \psi_{\downarrow \beta}(\vec{r}) \psi_{\uparrow \beta}(\vec{r}) \rangle.
\end{eqnarray}

\noindent Note that we have introduced a spatially uniform pairing order parameter, $\Delta_{\alpha}$. Even though unequal chemical potential for the atomic species $\uparrow$ and $\downarrow$ can favor a spatially modulated order parameter, we consider equal spin population of atoms in each atomic band. Proceeding with the Fourier transformation for the fermionic field using

\begin{eqnarray}
\psi_{\sigma \alpha}(\vec{r}) = \frac{1}{\sqrt{V}} \sum_k e^{i \vec{k} \cdot \vec{r}} \psi_{\sigma \alpha}(\vec{k})
\end{eqnarray}

\noindent with the condition

\begin{eqnarray}
\frac{1}{V} \int d^3\vec{r} e^{i (\vec{k} - \vec{k}^\prime) \cdot \vec{r}} = \delta_{k k^\prime},
\end{eqnarray}

\noindent followed by the usual Bogoliubov transformation, the mean-field Hamiltonian has the form~\cite{MOD5, MOD6}

\begin{eqnarray}
H = \sum_k \biggr[E^O_k (\beta^\dagger_{\uparrow k} \beta_{\uparrow k} + \beta^\dagger_{\downarrow -k} \beta_{\downarrow -k}) \\ \nonumber E^C_k (\xi^\dagger_{\uparrow k} \xi_{\uparrow k} + \xi^\dagger_{\downarrow -k} \xi_{\downarrow -k}) \biggr] + \Omega_0.
\end{eqnarray}

\noindent Here $\beta^{\dagger}_{\sigma k}$ and $\xi^{\dagger}_{\sigma k}$ represent the Bogoliubov quasiparticle creation operators for spin $\sigma$ in open ($\alpha =1$) and closed ($\alpha = 2$) atomic orbital bands. The excitation energies are defined as $E^\alpha_k = \sqrt{(\epsilon_k - \mu_\alpha)^2 + \Delta_\alpha^2}$, with $\epsilon_k = \hbar^2k^2/(2M)$. The zero-temperature grand potential is 

\begin{eqnarray}
\Omega_0 = \sum_{k, \alpha} (\epsilon_k - \mu_\alpha -E^\alpha_k) -\frac{2 |\Delta_{S}|^2}{V_S} -\frac{2 |\Delta_{T}|^2}{V_T}.
\end{eqnarray}

\noindent The combined superfluid order parameters $\Delta_{S}$ and $\Delta_{T}$ are defined as $\Delta_{T} = (\Delta_1 - \Delta_2)/2$ and $\Delta_{S} = (\Delta_1 + \Delta_2)/2$.

\noindent Due to the fact that the interaction potentials $V_{11}=V_{22}$ and $V_{12}=V_{21}$, we defined new interaction potentials,

\begin{eqnarray}
V_{11} + V_{12} = \frac{4\pi \hbar^2}{M}(2a_-) \equiv V_S
\\  \nonumber V_{11} - V_{12}= \frac{4\pi\hbar^2}{M}(2a_+)\equiv V_T.
\end{eqnarray}

\noindent They are related to the observed scattering lengths $a_S$ and $a_T$ through the renormalized interaction potentials~\cite{MOD5},

\begin{eqnarray}
\frac{1}{V_S} = \frac{M}{4\pi\hbar^2a_S} - \sum_k \frac{1}{2\epsilon_k}
\\ \nonumber
\frac{1}{V_T} = \frac{M}{4\pi\hbar^2a_T} - \sum_k \frac{1}{2\epsilon_k}.
\end{eqnarray}

\noindent These renormalizations will remove the divergences in the thermodynamic potential. Finally, we find the mean-field BCS grand potential,

\begin{eqnarray}
\Omega = -\frac{1}{\beta_T} \sum_{k \alpha} \ln[1 + e^{-\beta_T E_k^{\alpha}}] + \Omega_0,
\end{eqnarray}

\noindent where $\beta_T = 1/k_BT$ is the inverse thermal energy with Boltzmann's constant $k_B$ and temperature $T$.

\section{V. Self-consistent gap and number equations for intra-band pairing}

The self-consistent number equations and gap equations can be derived by minimizing the thermodynamic potential with respect to superfluid order parameters $\Delta_\alpha$ and chemical potentials $\mu_\alpha$. The number equation $n_{\alpha} = \sum_{\sigma} n_{\sigma \alpha}$ has the form,

\begin{eqnarray}
n_{\alpha} =\frac{1}{V}\sum_{k} \biggr(1- \frac{\epsilon_k - \mu_\alpha}{E_{k}^{\alpha}} \tanh(\beta_T E_{k}^{\alpha}/2) \biggr).
\end{eqnarray}

\noindent Defining two effective scattering lengths $a_{eff \pm}$, 

\begin{eqnarray}
\frac{1}{a_{eff \pm}} = \biggr( \frac{1}{a_S} \pm \frac{1}{a_T}\biggr)
\end{eqnarray}

\noindent the gap equation for the order parameter $\Delta_\alpha$ has the form,

\begin{eqnarray}
\frac{ \Delta_{\bar{\alpha}}}{\Delta_\alpha}\frac{M}{4 \pi \hbar^2 a_{eff -}} - \frac{M}{4 \pi \hbar^2 a_{eff +}} \\ \nonumber = \frac{2}{V} \sum_{k} \biggr(\frac{\tanh(\beta_T E_{k}^{\alpha})}{2E_{k}^{\alpha}} - \frac{1}{2 \epsilon_k}  \biggr).
\end{eqnarray}

\noindent Here $\bar{\alpha} = 2$ for $\alpha =1$ and vice versa. The newly defined effective scattering lengths $a_{eff \pm}$ are not independently controllable in experiments. For example, they were determined through $a_S$ and $a_T$ for the $^{173}$Yb system experimentally~\cite{XFR1, XFR2}. In our calculations, we use $a_S \approx 1900 a_0$ and $a_T \approx 200 a_0$, which are relevant for the $^{173}$Yb system, where $a_0 = 0.529 \AA$ is the Bohr radius. Using these values, we find the dimensionless effective scattering lengths $k_F a_{eff +} = 0.0629$ and $k_F a_{eff -} =-0.0777$, where the Fermi wave vector is defined through the atomic density $n$ as $k_F^{-1} = (3 \pi^2 n)^{-1/3}$. Experimentally, the BCS-BEC crossover is achieved by controlling the chemical potential through the magnetically controlled detuning $\delta$. The detuning dependence on the Feshbach scattering length $a_{FB}$ across the crossover is given by~\cite{MOD1, MOD7},

\begin{eqnarray}
a_{FB} = \frac{a_{so} - (a_{so}^2 -a_{s1}^2)\sqrt{m \delta/\hbar^2}}{1 - a_{so}\sqrt{m \delta/\hbar^2}}.
\end{eqnarray}

\begin{figure}
\includegraphics[width=0.9 \columnwidth]{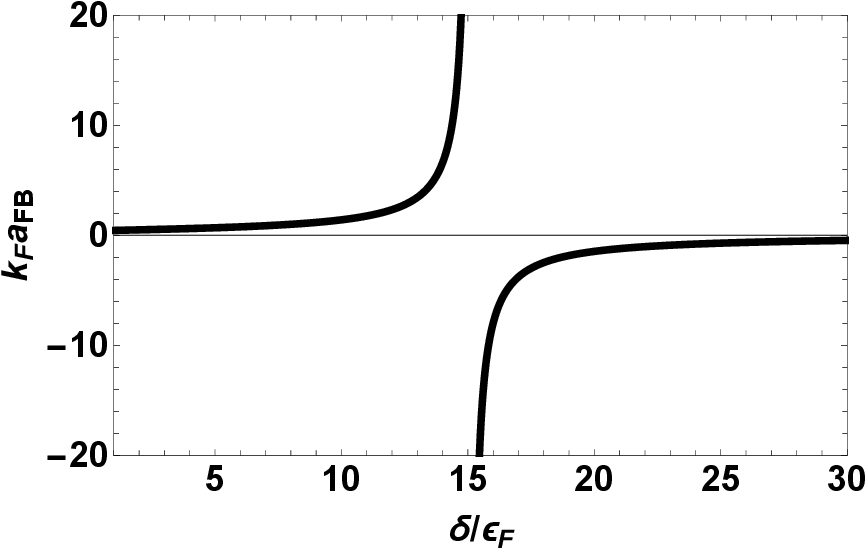}
\caption{Dimensionless effective s-wave scattering length as a function of dimensionless interchannel separation (detuning).}\label{PIC1}
\end{figure}

 \noindent The interchannel separation (detuning) dependence on the dimensionless s-wave scattering length $k_Fa_{FB}$ for the $^{173}$Yb system is shown in FIG.~\ref{PIC1}. Note that, experimentally, one can tune the scattering length by magnetically controlled detuning. Thus, experiments can access crossover from the weak coupling BCS regime [$k_fa_{FB} \rightarrow 0^{-}$ ($\delta \rightarrow \infty$)] to the weak coupling BEC regime [$k_fa_{FB} \rightarrow 0^{+}$ ($\delta \rightarrow 0$)] through the strongly coupled unitary regime [$k_fa_{FB} \rightarrow \pm \infty$ ($\delta \rightarrow \delta_{res}$)].

In order to solve the self-consistent equations as a function of effective scattering length $a_{FB}$, first, we convert the sums over wavevector $k$ into an integral $\frac{1}{V} \sum_k \rightarrow \int \frac{d^3k}{(2 \pi)^3}$, change the integration variable to $x_{\alpha} = \frac{\epsilon_k}{\sqrt{\mu_{\alpha}^2} + \Delta_{\alpha}^2}$, and rescale the energy units with Fermi energy $\epsilon_F = \hbar^2 k_F^2/(2 M)$. In these rescaled units, the length scales are measured in units of inverse Fermi wave vector  $k_F^{-1}$. They lead to the dimensionless number equation,

\begin{eqnarray} \label{dimenlessN}
n_\alpha = \frac{3}{4} (\tilde{\mu}_\alpha^2 + \tilde{\Delta}_\alpha^2)^{3/4} I_{n_\alpha}(z_\alpha, T),
\end{eqnarray}

\noindent where the atomic density in the $\alpha$ band is measured in units of $k_F^3/(3 \pi^2)$. The tilde notations are dimensionless quantities scaled with Fermi energy and $z_{\alpha} = -\frac{\tilde{\mu}_{\alpha}}{\sqrt{\tilde{\mu}_{\alpha}^2 + \tilde{\Delta}_{\alpha}^2}}$. The dimensionless integral $I_{n_{\alpha}}(z_\alpha, T)$ is defined as

\begin{eqnarray}
I_{n_{\alpha}}(z_\alpha, T) = \int^\infty_0 \sqrt{x} \biggr\{1- \frac{x + z_{\alpha}}{E_{\alpha}(z_\alpha)} 
\\ \nonumber
\times \tanh[\tilde{\beta}_T E_{\alpha}(z_\alpha)/2]   \biggr\} dx.
\end{eqnarray}

\noindent The dimensionless gap equation has the form,

\begin{eqnarray}
\frac{ \Delta_{\bar{\alpha}}}{\Delta_\alpha}\frac{1}{k_F a_{eff -}} - \frac{1}{k_F a_{eff +}}  \\ \nonumber = \frac{2}{\pi} \biggr[(\tilde{\mu}_{\alpha}^2 + (\tilde{\Delta}_{\alpha}^2)^{1/4} I_{\Delta_{\alpha}}(z_{\alpha}, T) \biggr],
\end{eqnarray}

\noindent where $E_{\alpha}(z_\alpha) = \sqrt{(x + z_{\alpha})^2 - z_{\alpha}^2 +1}$. The dimensionless integral $I_{\Delta_{\alpha}}(z_\alpha, T)$ is defined as,

\begin{eqnarray}
I_{\Delta_{\alpha}}(z_\alpha, T) = \int^\infty_0 \sqrt{x} \biggr\{\frac{\tanh[\tilde{\beta}_T E_{\alpha}(z_\alpha)/2]}{E_{\alpha}(z_\alpha)}  \\ \nonumber
- \frac{1}{x}\biggr\} dx.
\end{eqnarray}

\noindent Here we have used the fact that $z_{\alpha}^2 + \tilde{\Delta}_{\alpha}^2 = 1$ to simplify our equations. We use these equations  to find exact BCS analytical solutions at the zero-temperature limit in the next section.

\section{VI. Zero temperature static properties from exact BCS analytical results}

At zero temperature, the integrals represented in Eqs. (18) and (20) have analytical forms~\cite{MM},

\begin{eqnarray}
I_{n_{\alpha}}(z_\alpha, T=0) = z_\alpha \pi P_{1/2}(z_\alpha) - \pi P_{3/2}(z_\alpha)
\end{eqnarray}

\noindent and 

\begin{eqnarray}
I_{\Delta_{\alpha}}(z_\alpha, T = 0) = - \pi P_{1/2}(z_\alpha),
\end{eqnarray}

\noindent where $P_n(z)$ is the Legendre polynomial of order $n$. Using a series expansion, then integrating term by term, and combining the integrated terms again, these zero-temperature analytical equations were derived for single-band superfluids by Obeso-Jureidini and Romero-Rochin~\cite{Rochin}. Our zero-temperature analytical expressions for the two-band case can be considered a generalization of those single-band results. 

Finally, we combine the zero-temperature number and gap equations to arrive at the final analytical forms, 

\begin{eqnarray} \label{dimenlessN}
n_\alpha = \frac{3 \pi}{4} (\tilde{\mu}_\alpha^2 + \tilde{\Delta}_\alpha^2)^{3/4} \biggr(\tilde{z}_{\alpha} P_{1/2}(\tilde{z}_{\alpha}) - P_{3/2}(\tilde{z}_{\alpha}) \biggr),
\end{eqnarray}

\noindent

\begin{eqnarray}
\frac{ \Delta_{\bar{\alpha}}}{\Delta_\alpha}\frac{1}{k_F a_{eff -}} - \frac{1}{k_F a_{eff +}} =2(\tilde{\mu}_{\alpha}^2 + \tilde{\Delta}_{\alpha}^2)^{1/4} P_{1/2}(\tilde{z}_{\alpha}),
\end{eqnarray}

\noindent where thermodynamic variables denoted by tilde notations are dimensionless with energies scaled by the Fermi energy and length scaled by the inverse Fermi wave vector. As the density is also scaled by $k_F^3/(3 \pi^2)$, we have $n_1 + n_2  =1$.

\begin{figure}
\includegraphics[width=0.9 \columnwidth]{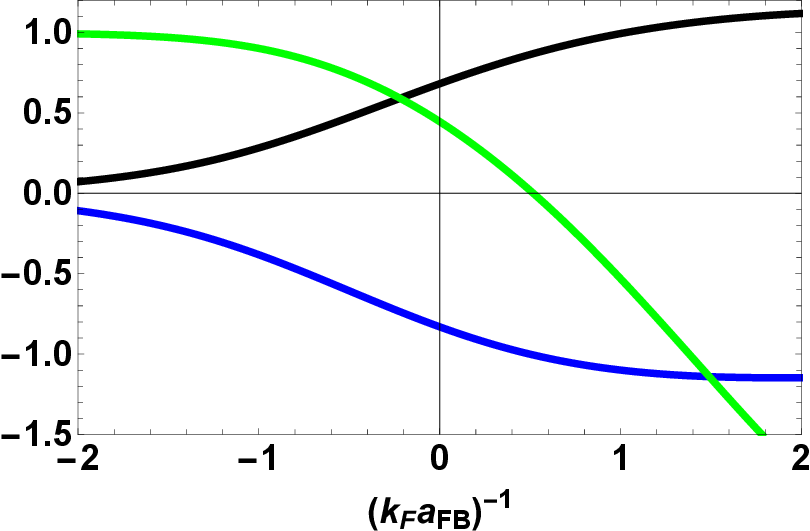}
\caption{(color online) Zero-temperature superfluid order parameters $\Delta_1/\epsilon_F$ [black (upper)] and $\Delta_2/\epsilon_F$ [blue (lower)] and open-channel chemical potential $\mu_1/\epsilon_F$ [green (light gray)] as a function of the dimensionless inverse scattering length $(k_Fa_{FB})^{-1}$}\label{PIC2}
\end{figure}

\begin{figure}
\includegraphics[width=0.9 \columnwidth]{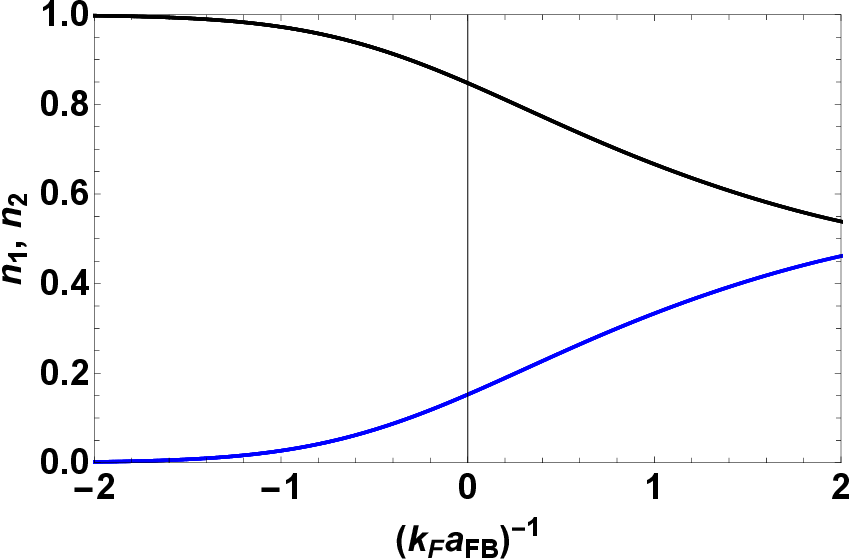}
\caption{(color online) Zero-temperature atomic density variation across the crossover region for the open channel ($n_1$) and closed channel ($n_2$). Dimensionless densities in the open channel [$n_1$: black (upper)] and the closed channel [$n_2$: blue (lower)] are scaled with $k_F^3/(3 \pi^2)$.}\label{PIC3}
\end{figure}

The gap equations have two types of solutions for the superfluid order parameters~\cite{MOD3, MOD5, MOD6, MOD14, MOD16}. One is the \emph{in-phase} solution where sign($\Delta_1$) = sign($\Delta_2$) which corresponds to a deep bound state with a global minimum of thermodynamic potential. The other is the \emph{out-of-phase} solution where sign($\Delta_1$) = -sign($\Delta_2$) which corresponds to a shallow bound state with an exited saddle point of thermodynamic potential. The \emph{in-phase} deep bound-state solution requires a large negative chemical potential for all values of detuning. Thus, we restrict ourselves to  the \emph{out-of-phase} solution responsible for the recently observed crossover near the OFR of  the $^{173}$Yb system. For given values of $\delta$ across the BCS-BEC crossover, we calculate the $a_{FB}$ and deduce the superfluid order parameters and chemical potentials from our zero-temperature exact BCS solutions. In FIG~\ref{PIC2}, we plot the zero-temperature superfluid order parameters $\Delta_\alpha$ and chemical potential $\mu_1$ as a function of effective inverse scattering length $(k_fa_{FB})^{-1}$ for the both $\alpha = 1$, and $\alpha = 2$ bands. In the entire crossover region, the superfluid order parameter for the closed channel ($\alpha =2$) is comparable to that of the open channel ($\alpha =1$) and opposite in sign, regardless of the atomic density in the closed channel. The zero-temperature atomic densities in open and closed channels across the crossover region are plotted in FIG.~\ref{PIC3}. Note that the densities are scaled so that the sum of the atomic densities in the two bands is equal to unity. As expected, while the magnitudes of the superfluid order parameters are small in the BCS regime where $k_fa_{FB} \rightarrow 0^{-}$ ($\delta \rightarrow \infty$), they are large in the BEC regime where $k_fa_{FB} \rightarrow 0^{+}$ ($\delta \rightarrow 0$). Note the polarization $P = (n_1 - n_2)/(n_1+n_2)$ varies from unity to zero across the BCS-BEC crossover. It is interesting to see that the superfluid order parameter for the closed channel is comparable to that of the open channel even in the weak BCS limit where the closed-channel atomic density $n_2 \rightarrow 0$. This is not surprising because the superfluid order parameter is related to the binding energy of a Cooper pair but not to the number of Bose-condensed Cooper pairs, which is related to the atomic density. Our analytical results for the order parameters and the chemical potential in FIG.~\ref{PIC2} are consistent with numerical results presented in Refs.~\cite{MOD5, MOD6}.

\subsection{Pair-correlation length and Phase-coherence length across the BCS-BEC crossover}

In addition, we calculate the pair correlation length $\xi_{pair}$ and the phase coherence length for the both bands $\xi_{phase}$~\cite{CLFCL}:

\begin{eqnarray}
\xi_{pair, \alpha}^2 = \frac{\hbar^4}{m^2} \frac{\sum_k \frac{k^2 (\epsilon_k - \mu_\alpha)^2}{E_{k \alpha}^6}}{\sum_k \frac{1}{E_{k \alpha}^2}},
\end{eqnarray}

\begin{eqnarray}
\xi_{phase, \alpha}^2 = \frac{\hbar^2}{8 m} \times \\ \nonumber \frac{\sum_k \biggr\{ \frac{(\epsilon_k - \mu_\alpha)^3 -2 (\epsilon_k - \mu_\alpha) \Delta_\alpha^2}{E_{k \alpha}^5} + \frac{5 \hbar^2k^2 \Delta_\alpha^2 (\epsilon_k - \mu_\alpha)^2}{3 mE_{k \alpha}^7}\biggr\}}{\sum_k \frac{\Delta_{\alpha}^2}{E_{k \alpha}^3}}.
\end{eqnarray}

\noindent Converting the sums into integrals and changing the integration variable as before, we find the dimensionless form of the correlation and coherence lengths,

\begin{eqnarray}
\xi_{pair, \alpha}^2 k_F^2 = \frac{4}{\sqrt{\tilde{\mu_\alpha}^2 + \tilde{\Delta_\alpha}^2}} \frac{I_{pairA}(z_\alpha)}{I_{pairB}(z_\alpha)},
\end{eqnarray}

\begin{eqnarray}
\xi_{phase, \alpha}^2 k_F^2 = \frac{1}{4 \sqrt{\tilde{\mu_\alpha}^2 + \tilde{\Delta_\alpha}^2}} \frac{I_{phaseA}(z_\alpha)}{I_{phaseB}(z_\alpha)},
\end{eqnarray}

\noindent where we defined dimensionless integrals as,

\begin{eqnarray}
I_{pairA}(z_\alpha) = \int_0^\infty \frac{x^{3/2} (x + z_\alpha)^2 dx}{[(x + z_\alpha)^2 - z_\alpha^2 +1]^3}    \\ \nonumber
I_{pairB}(z_\alpha) = \int_0^\infty \frac{x^{1/2} dx }{[(x + z_\alpha)^2 - z_\alpha^2 +1]}  \\ \nonumber
I_{phaseA}(z_\alpha) = \frac{10 (1-z_\alpha^2)}{3} \int_0^\infty \frac{x^{3/2} (x+z_\alpha)^2 dx }{[(x + z_\alpha)^2 - z_\alpha^2 +1]^{7/2}} +
\\ \nonumber\int_0^\infty \frac{x^{1/2} [(x + z_\alpha)^3 -2 (x + z_\alpha) (1-z_\alpha^2)]dx}{[(x + z_\alpha)^2 - z_\alpha^2 +1]^{5/2}} \\ \nonumber 
I_{phaseB}(z_\alpha) =  (1-z_\alpha^2) \int_0^\infty \frac{x^{1/2} dx }{[(x + z_\alpha)^2 - z_\alpha^2 +1]^{3/2}}.
\end{eqnarray}

\begin{figure}
\includegraphics[width=0.9 \columnwidth]{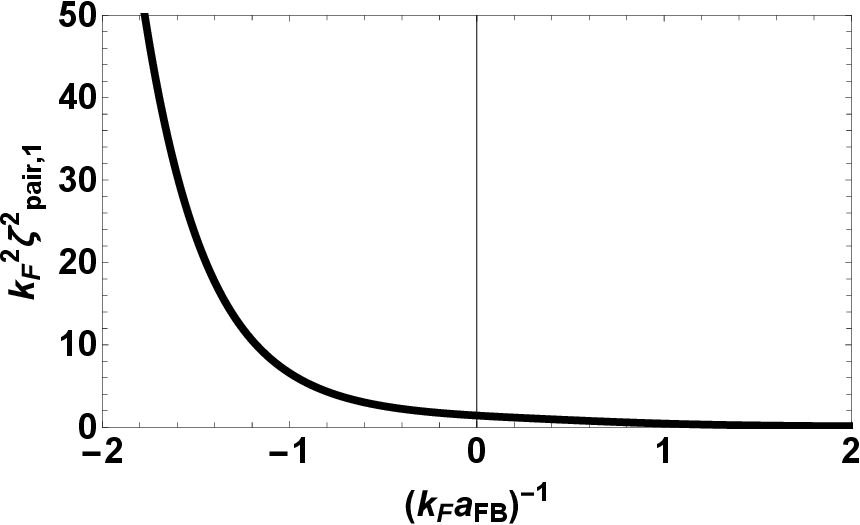}
\caption{Dimensionless pair correlation length $ \xi_{pair, 1}$ as a function of dimensinless inverse effective scattering length for the open-channel ($\alpha =1$) atoms.}\label{PIC4}
\end{figure}

\begin{figure}
\includegraphics[width=0.9 \columnwidth]{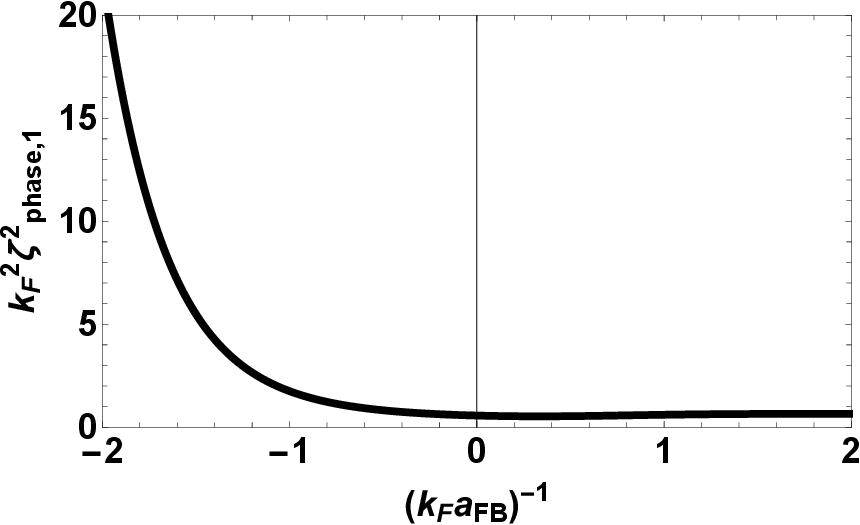}
\caption{Dimensionless phase coherence length $\xi_{phase,1}$ as a function of dimensionless inverse effective scattering length for the open-channel ($\alpha =1$) atoms.}\label{PIC5}
\end{figure}

As a demonstration, both the pair-correlation length and phase-coherence length for the open-channel atoms are plotted as a function of effective inverse scattering length in FIGS.~\ref{PIC4} and \ref{PIC5}. As seen from the FIGS.~\ref{PIC4} and \ref{PIC5}, both lengths are larger in the effective BCS limit and smaller in the effective BEC limit. The pair-correlation length is the characteristic length for the pair correlation and can be considered to be the range of effective interaction. The phase-coherence length is the length associated with the spatial fluctuation of the superfluid order parameter. It can be considered to be the effective pair size of the two paired atoms. Therefore, both pair-correlation and phase-coherence lengths show the bosonic and fermionic natures in the effective BCS and BEC limits, respectively. These coherence lengths have links to experimental observables. For example, the phase-coherence length can be considered to be the length scale below which the amplitude and phase of the collective Higgs mode is significantly coupled~\cite{HIGGS}.

\subsection{Atomic density profiles across the crossover region}

We assume that the atomic cloud is confined by a harmonic-oscillator
trapping potential given by $V (r) = \frac{1}{2} M [\omega_\perp^2 (r_1^2 + r_2^2) + \omega_z^2 r_3^2]$, where $\omega_\perp$ and $\omega_z$ are transverse and axial trapping frequencies. Here $r_i = (x, y, z)$ for $i = 1 ,2, 3$ are the Cartesian coordinates of the position vector $\vec{r}$. Introducing a dimensionless length $\tilde{r_i} = (r_i/\xi)(\omega_i/\omega)$ with $\omega = (\omega_\perp^2 \omega_z)^{1/3}$ and $\xi = \sqrt{\hbar/(2 \pi M \omega)}$, we can write the trapping potential in the spherically symmetric form, $V(\tilde{r}) = \hbar \omega \tilde{r}^2/(4 \pi)$.  Assuming that the atomic density is locally homogeneous, we use the local-density approximation (Thomas-Fermi approximation) by taking the local chemical potential $\mu_\alpha = \mu_{\alpha, 0} - V(\tilde{r})$ to calculate the atomic density profiles. Here the central chemical potential $\mu_{\alpha, 0}$ is determined by fixing the total number of atoms $N = \sum_\alpha N_\alpha$ in the trap using,

\begin{eqnarray}
N_\alpha = 4 \pi \int_0^{\tilde{r}_{0}} \tilde{r}^2 n_{\alpha}(\tilde{r}) d\tilde{r},
\end{eqnarray}

\noindent where $\tilde{r}_{0}$ is the edge of the trap. We use the total number of atoms $N = 2 \times 10^5$ with the Fermi energy $\epsilon_F = \hbar \omega (6 N)^{1/3}$ to find the density profiles $n_\alpha (\tilde{r})$ within the local-density approximation where the spatial dependence, $\tilde{r}$ on the density is included through the local chemical potential.

\begin{figure}
\includegraphics[width=0.9 \columnwidth]{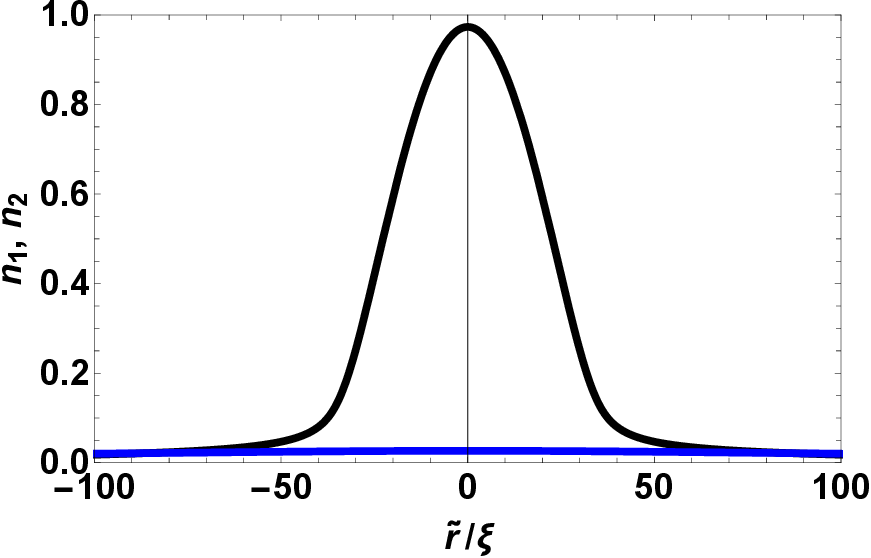}
\caption{(color online) Dimensionless density profile of the open-channel [black (upper)] and closed-channel [blue (lower)] atoms in the BCS regime where $(k_{F} a_{FB})^{-1} = -1.0$. The densities are scaled with the factor $k_F^3/(3 \pi^2)$, and the effective radial coordinate $\tilde{r}$ is scaled with $\xi =\sqrt{\hbar/(2 \pi M \omega)}$.}\label{PIC6}
\end{figure}

\begin{figure}
\includegraphics[width=0.9 \columnwidth]{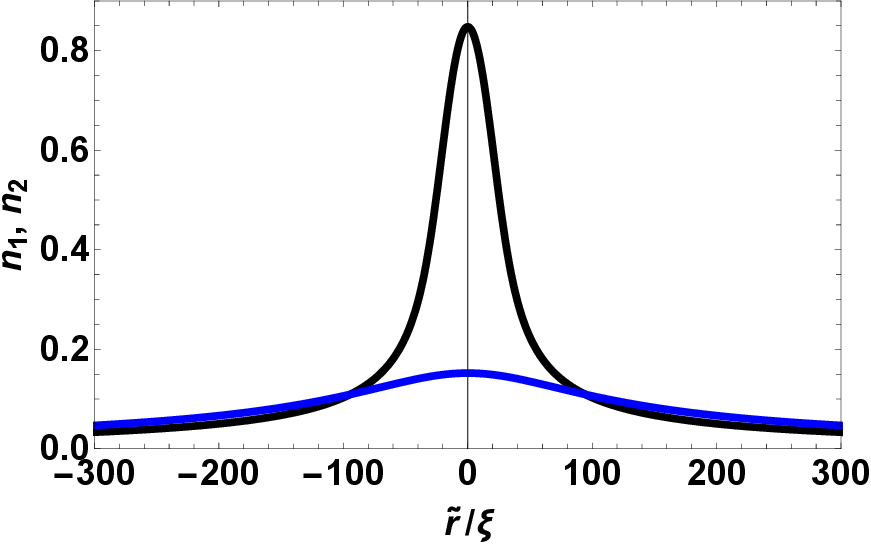}
\caption{(color online) Dimensionless density profile of the open channel [black (upper)] and closed channel [blue (lower)] atoms in the unitary regime where $(k_{F} a_{FB})^{-1} = 0.0$. Densities are scaled with the factor $k_F^3/(3 \pi^2)$, and the effective radial coordinate $\tilde{r}$ is scaled with $\xi =\sqrt{\hbar/(2 \pi M \omega)}$.}\label{PIC7}
\end{figure}

\begin{figure}
\includegraphics[width=0.9 \columnwidth]{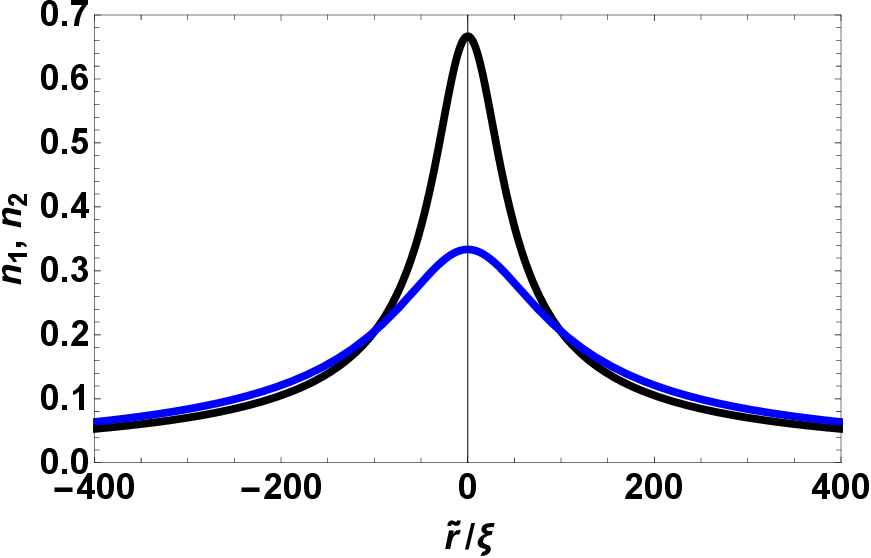}
\caption{(color online) Dimensionless density profile of the open-channel [black (upper)] and closed-channel [blue (lower)] atoms in the BEC regime where $(k_{F} a_{FB})^{-1} = 1.0$. The densities are scaled with the factor $k_F^3/(3 \pi^2)$, and the effective radial coordinate $\tilde{r}$ is scaled with $\xi =\sqrt{\hbar/(2 \pi M \omega)}$.}\label{PIC8}
\end{figure}

The calculated Thomas-Fermi density profiles in the BCS, unitary, and BEC regimes are shown in FIG~\ref{PIC6}, FIG~\ref{PIC7}, and FIG~\ref{PIC8}. For a better comparison, we fixed the total number of particles and the trapping frequency so that total peak density at the center of the trap is unity. Consistent with the previous homogeneous density shown in FIG~\ref{PIC2}, the closed-channel density in the BCS regime is very small where the spatial variation of the open-channel density profile is narrow. However, as the closed channel populates with excited atoms as one goes from the BCS $\rightarrow$ unitary $\rightarrow$ BEC regime, the density profiles spread spatially.

\subsection{Ground-state energy density and Tan's contact density across the crossover}

Due to the analytical nature of the number equations and the gap equations, the total ground-state energy density $e_0 = E_0/(\epsilon_F V)$ also has the analytical form, 

\begin{eqnarray}
e_0 = -\frac{1}{5} \biggr[ \frac{\tilde{\Delta}_1^2 + \tilde{\Delta}_2^2}{k_F a_{eff +}}- \frac{2 \tilde{\Delta}_1 \tilde{\Delta}_2}{k_F a_{eff -}} - 3 \sum_\alpha \tilde{\mu}_{\alpha} n_{\alpha} \biggr].
\end{eqnarray}

\begin{figure}
\includegraphics[width=0.9 \columnwidth]{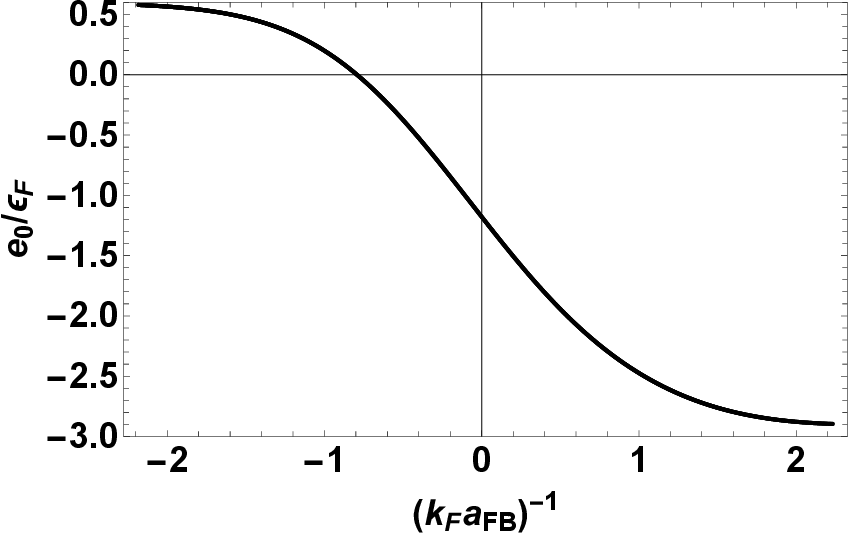}
\caption{Dimensionless ground state energy density variation across the crossover region for the two-band atomic system. The horizontal axis represents the dimensionless inverse scattering length.}\label{PIC9}
\end{figure}

\noindent A similar expression for single-band systems is found in Refs.~\cite{Rochin, CLFCL, en1, en2}, and our $e_0$ can be considered a generalization of those for the two-band systems. The calculation of the ground-state energy, 

\begin{eqnarray}
E_0 = \frac{\langle \psi_0|H| \psi_0 \rangle}{\langle \psi_0|\psi_0 \rangle}, 
\end{eqnarray}

\noindent does not require renormalization because the chemical potentials and superfluid order parameters are calculated self-consistently from renormalized number equations and gap equations~\cite{en1}. In this ground-state energy density expression, the molecular energy or the binding energy of Cooper pairs is already included. The calculated total energy density across the BCS-BEC crossover region is presented in FIG.~\ref{PIC9}. 

Due to the fact that the range of the interaction is much smaller than the interparticle distance in cold-atom experiments, a set of \emph{universal} relations was derived by Tan~\cite{tan1, tan2, tan3}. These exact universal relations for Fermi gases across the entire BCS-BEC crossover region connect microscopic properties to the thermodynamic quantities. The connection is made through a single quantity known as the \emph{Tan's contact} $C$, 

\begin{eqnarray}
C = \lim_{k\to \infty} k^4 n(k),
\end{eqnarray}

\noindent which is the high-momentum ($k$) tail of the momentum distribution $n(k)$. Remarkably, all the properties of the short-range interacting many-body system are carried by Tan's contact. One of Tan's relations, the adiabatic sweep theorem, connects Tan's contact density ($c =C/V$) to the change in energy density due to the adiabatic change in the scattering length $a$~\cite{theja1},

\begin{eqnarray}
c = \frac{4 \pi M a^2}{\hbar^2} \biggr (\frac{\partial e_{gs}}{\partial a}\biggr)_{\mu, T}.
\end{eqnarray}

\begin{figure}
\includegraphics[width=0.9 \columnwidth]{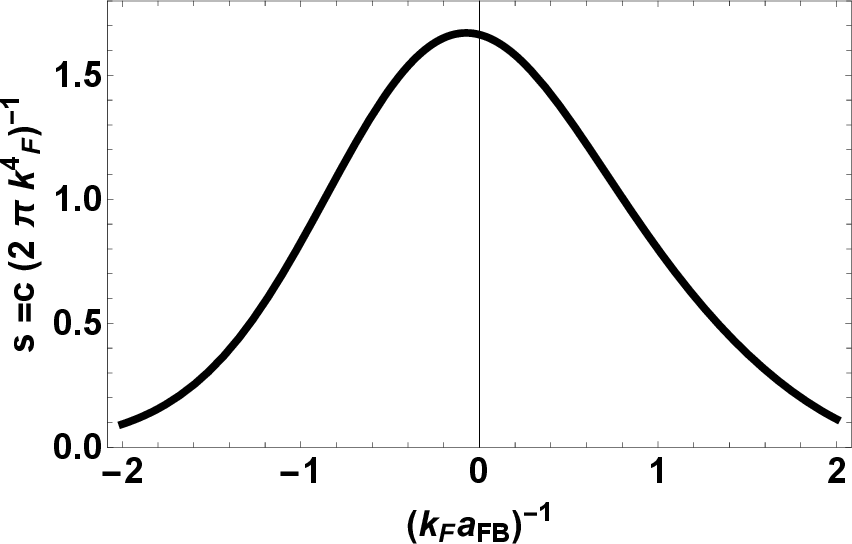}
\caption{Dimensionless Tan's contact density variation across the crossover region for the two-band OFR atomic system. This single contact density for the entire atomic system is calculated from the adiabatic theorem.}\label{PIC10}
\end{figure}

\noindent We generalize the definition for our OFR system by $a \rightarrow a_{FB}$ and $e_{gs} \rightarrow e_0$. By defining a dimensionless Tan's contact density $s = c/k_F^4$, we find,

\begin{eqnarray}
s = -2 \pi \frac{\partial e_0}{\partial \chi}
\end{eqnarray}

\noindent where $\chi = (k_F a_{FB})^{-1}$. Using the energy density presented in FIG.~\ref{PIC9}, we calculate the dimensionless Tan's contact density $s$ across the BCS-BEC crossover region and the result is presented in FIG.~\ref{PIC10}. As can be seen from FIG.~\ref{PIC10}, the Tan's contact density increases as one goes from the weak BCS limit ($k_F a_{FB} \rightarrow 0^{-}$) to the unitary limit ($k_F a_{FB} \rightarrow \infty$). This is consistent with single-band superfluid systems~\cite{tanE1}. On the opposite side, Tan's contact density also increases as one goes from the weak BEC limit ($k_F a_{FB} \rightarrow 0^{+}$) to the unitary limit ($k_F a_{FB} \rightarrow \infty$). This behavior of the single-contact density in the BEC regime is \emph{opposite} to that of single band superfluid systems. This behavior of the OFR systems can be understood from Tan's adiabatic sweep theorem~\cite{tanE2, tanE3}. The adiabatic theorem links the short-range behavior to the contact; thus, the contact is a measure of the probability of two opposite spins being close together. As shown in FIG.~\ref{PIC3}, the density in the closed channel $n_2$ is very small in the BCS regime, and the closed-channel atoms mostly populate in the BEC regime. As the closed-channel atoms populate, the atoms spatially spread broadly in the BEC regime, as seen from FIG.~\ref{PIC6}, \ref{PIC7}, and \ref{PIC8}. Therefore, on average, the probability of two opposite spins being close together is smaller in the BEC regime for OFR systems.  

The single Tan's contact density shown in FIG. 10 is derived as the thermodynamic quantity conjugate to the effective inverse scattering length. Alternatively, one can define two contact parameters $c_\alpha$ as the high-momentum tail of atomic densities in two bands~\cite{dcd1}. The Fourier transform of the density of the $\alpha$ band given in Eq.~(23), has the form

\begin{eqnarray}
n_\alpha (k)  = \frac{1}{2} \biggr(1- \frac{\epsilon_k -\mu_\alpha}{E^\alpha_k} \biggr).
\end{eqnarray}

\noindent Expanding this for large \emph{k}, 

\begin{eqnarray}
n_\alpha (k)  \approx \frac{\Delta_\alpha^2}{4 (\epsilon_k -\mu_\alpha)^2}  \approx \frac{m^2 \Delta_\alpha^2}{\hbar^4 k^4},
\end{eqnarray}

\noindent we find 

\begin{eqnarray}
c_\alpha =  \frac{m^2 \Delta_\alpha^2}{\hbar^4}.
\end{eqnarray}

\begin{figure}
\includegraphics[width=0.9 \columnwidth]{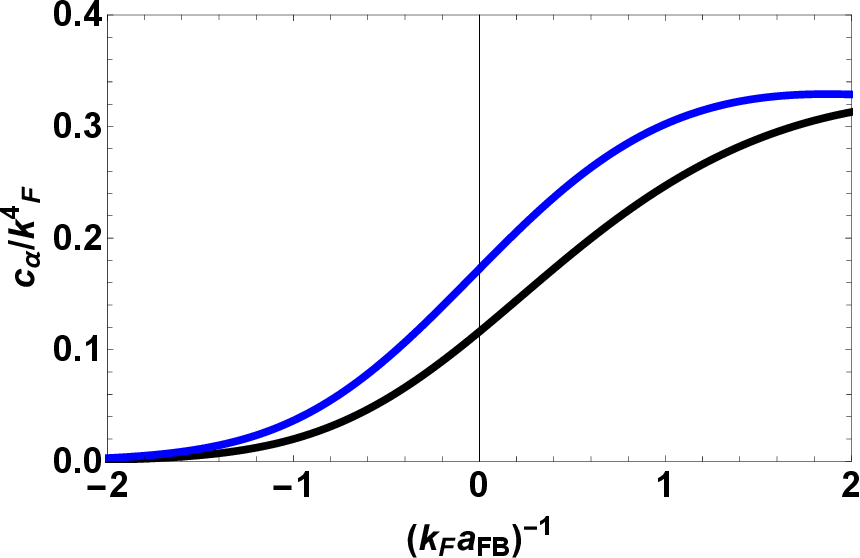}
\caption{Dimensionless Tan's contact density $c_\alpha$ for atoms in the open channel [$\alpha =1$; black (lower)] and the closed channel [$\alpha =2$; blue (upper)] across the BCS-BEC crossover region. These double-contact densities are calculated from the high-momentum tail of two atomic densities.}\label{PIC11}
\end{figure}

\noindent This double-contact density for the two-band system is shown in FIG.~\ref{PIC11}. Unlike the single-contact density shown in FIG.~\ref{PIC10}, the behavior of the double contact density across the BCS-BEC crossover region is consistent with that of the contact density of single-band superfluidity~\cite{theja1, tanE1, dcd2}. Note that the behavior of combined single-contact density shown in FIG.~\ref{PIC10} is different from that of the double-contact density shown in FIG.~\ref{PIC11} in the BEC regime. This difference is due to the fact that single-contact density incorporates fermionic band effects. While single-contact density is related to the combined atomic densities in both bands, the double contact-density is related to the atomic densities in individual bands; thus, the double-contact density behaves like that of single-band superfluidity.

\section{VII. Conclusions and summary}

We studied a two-band superfluid system of alkali-earth-like atoms in the presence of orbital Feshbach resonance in which one can control both intra- and inter-orbital interactions. We used a simple mean-field theory to decouple the attractive interactions between atoms and derive self-consistent equations for the superfluidity. At zero-temperature, we derived exact analytical solutions for the BCS equations and used these results to predict various static properties. Our exact BCS analytical results can be used in connection with experiments to test the validity of mean-field theory. Further, they can be used as bench mark calculations for heavy numerical calculations. Our findings and prediction of static properties can be experimentally tested with already available experimental tools using general species of fermionic alkaline-earth atoms, such as $^{173}$Yb with nuclear spin $I = 5/2$, $^{171}$Yb with nuclear spin $I = 1/2$, and $^{87}$Sr with nuclear spin $I = 9/2$~\cite{XFR1, AEM1, AEM2, AEM3, AEM4}.

\section{VIII. ACKNOWLEDGMENTS}

We are grateful and acknowledge the partial support from Augusta University's Provost's Student Research Program (SRP) and Center for Undergraduate Research and Scholarship (CURS) through travel grants.

\end{document}